\documentclass[runningheads]{llncs}

\usepackage{float}
\usepackage{listings}
\usepackage{graphicx}

\title{The CEKG: A Tool for Constructing Event Graphs in the Care Pathways of Multi-Morbid Patients}
\author{Milad Naeimaei Aali\inst{1} \and
Felix Mannhardt\inst{2}\and
Pieter Jelle Toussaint\inst{1}}
\authorrunning{M. Naeimaei Aali}
\institute {
Norwegian University of Science and Technology, Trondheim, Norway 
\and Eindhoven University of Technology, Eindhoven, Netherlands
}

\begin{document}

\maketitle

\begin{abstract}
One of the challenges in healthcare processes, especially those related to multi-morbid patients who suffer from multiple disorders simultaneously, is not connecting the disorders in patients to process events and not linking events' activities to globally accepted terminology. Addressing this challenge introduces a new entity to the clinical process. On the other hand, it facilitates that the process is interpretable and analyzable across different healthcare systems. This paper aims to introduce a tool named CEKG that uses event logs, diagnosis data, ICD-10, SNOMED-CT, and mapping functions to satisfy these challenges by constructing event graphs for multi-morbid patients' care pathways automatically.


\end{abstract}

\section{Introduction}

For both ethical reasons and economic progress, it is essential to foster a society with healthy people. Achieving this vision necessitates the presence of robust and effective healthcare services. These services are crucial to ensure timely medical treatment and preventative care for all. Patients with multi-morbidity in particular need such care. This patient group, who have multiple chronic conditions at the same time~\cite{marengoni2011aging}, is expanding due to socio-economic deprivation and an aging population~\cite{marengoni2011aging}, and they require coordinated care from various specialties. They also demand significantly more resources due to the complexity of their conditions. Therefore, enhancing healthcare services for such patients can be a cornerstone of achieving truly effective healthcare for all. One of the approaches for enhancing healthcare service for these multi-morbid patients is enhancing the clinical process they are subject to.

A clinical process or care pathway outlines the events involved in diagnosing, treating, managing, and following up with patients~\cite{munoz2022process}. It can be considered a type of business process and, consequently, techniques like process mining may be used to improve a multi-morbid patient's clinical process. Still, there are a lot of challenges when applying process mining methods to the care pathways of multi-morbid patients, which often spread several caregivers in multiple organizations and involve the simultaneous treatment of multiple conditions. Among these challenges are connecting emerging entities to events and linking relevant terminology to them, Addressing both of these challenges may significantly enhance the delivery of care paths~\cite{naeimaei2023clinical}.

\begin{itemize}
    \item \textbf{Connecting clinical entities.} Emerging clinical entities are clinical attributes that are not connected directly to events but can potentially be used as new entities. For example, multi-morbid patient disorders are connected to the patient entity but not attached to events. Multi-morbid patients have many different (sometimes emerging) disorders, which we see as entities connected to the patient. By connecting events to the relevant disorders, thereby getting a multi-entity event data, we can better query the event data of a patient to find relevant insights\cite{naeimaei2023clinical}. 

    \item \textbf{Linking terminology.} Standardized clinical coding and nomenclature systems provide useful terminology that are often not linked to the clinical process's activities and entities. For example, sources like Systematized Nomenclature of Medicine Clinical Terms (SNOMED CT) [7], International Classification of Diseases Clinical Modification (ICD CM) [8], and diagnosis-related groups (DRG) that store event activities and entities terminology in a standardized way are not linked directly to event activities and entities. By aligning terminology with activities and events, we can standardize clinical processes, enabling global interoperability for patient diagnoses and event activities. This may also allow for various levels of abstraction and standardized categorization, ensuring a more organized and segmented process\cite{naeimaei2023clinical}.
\end{itemize}

In this paper we introduce a tool for the Clinical Event Knowledge Graph (CEKG) framework presented in our previous work~\cite{naeimaei2023clinical}, which addresses the two mentioned challenges. We developed this tool to support constructing CEKGs for enhancing the process analysis of multi-morbid patient care pathways. The tool utilizes inputs such as low-dimensional clinical event logs, diagnostic data (indicating each patient's disorders), ICD codes, and SNOMED CT terminology. It allows to map the different inputs through constrained node mappings, which are functions derived from various sources, including empirical data, domain expertise, professional insights, and documentation. 

In Section~\ref{sec:tool}, we describe the innovations of this tool and in Sect.~\ref{sec:case} we show its application in a case study. 

\section{The Overview of the Tool}
\label{sec:tool}


Since the tool needs to support terminologies such as SNOMED-CT and ICD-10 as parts of its inputs a graph database was chosen for storage since it supports a linked-data structure for these terminologies. Furthermore, the need for path-based traversal of data makes the graph database an ideal choice. Additionally, the tool requires the storage of entity attributes and other semantic patient data, further reinforcing the suitability of a graph database for these functions. The CEKG was proposed using the Labelled Property Graph Model. However, a challenge in creating a CEKG using Neo4J is integrating data from different sources (SNOMED, hospital information systems, etc.), which requires using several complex Cypher Query Language (CQL) queries manually. ~\cite{swevels2023object} introduced an open-source Python library for exploring graph-based, object-centric process discovery, but this approach requires the deployment of clinical data and terminology.

In addressing usability challenges, process mining tools frequently exhibit operational complexities, particularly when applied by healthcare professionals managing patients with multi-morbid conditions. To address these difficulties, the CEKG tool incorporates a user interface that is designed, enabling users to rapidly assimilate its functionalities from the initial steps. Furthermore, the tool generates outputs in the LPG format and employs the Graphviz library for visualization purposes.

\begin{figure}[tb]
    \centering
    \includegraphics[page=1, scale=0.47]{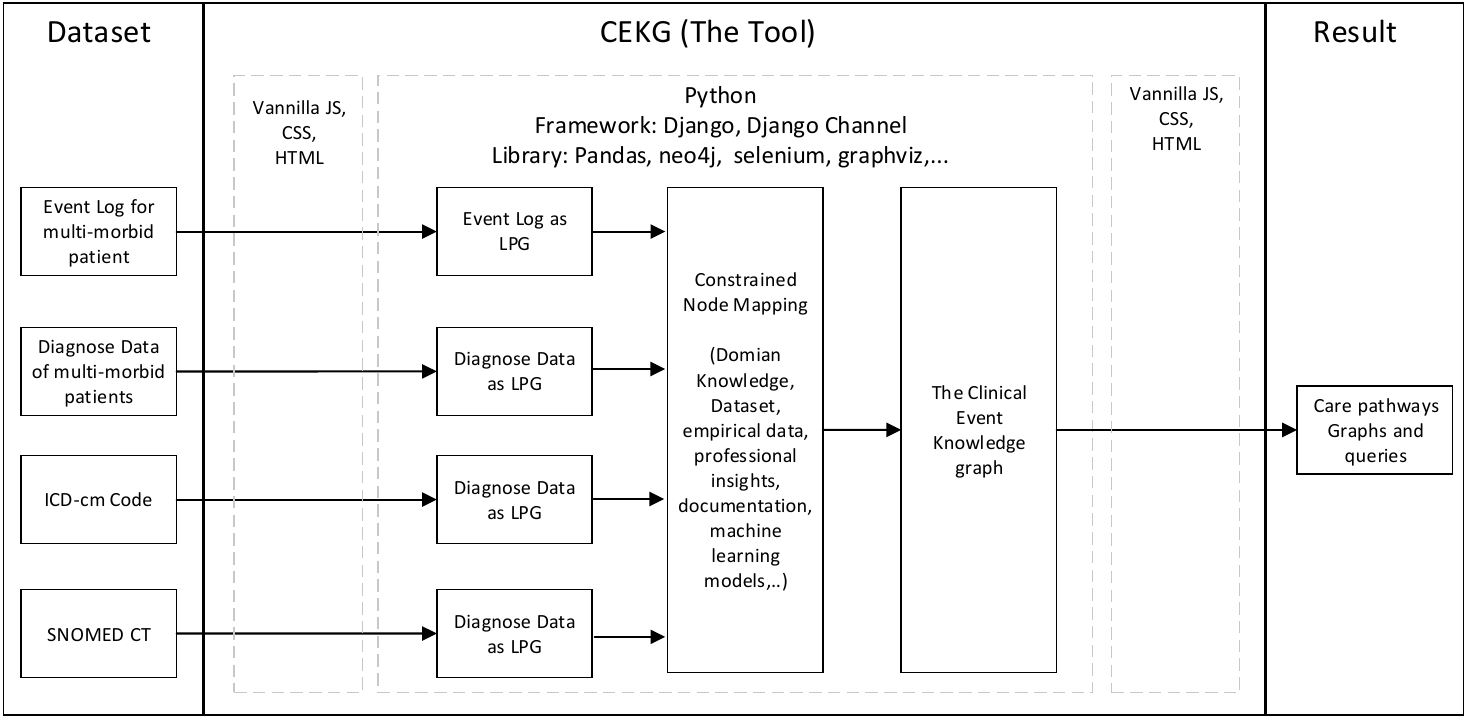}\newline
    \caption{The tool was developed using Python with Django and Django Channels as the backend framework, supplemented by libraries including Pandas, Neo4j, Selenium, and Graphviz. The frontend was created using vanilla JavaScript, HTML, and CSS.}
    \label{toolFirstPageFigure}  
\end{figure}

For the implementation of the tool, as illustrated in Fig.~\ref{toolFirstPageFigure}, Python with Django and Django Channels was used as the backend framework along with several libraries such as Pandas, Neo4j, and Graphviz. For frontend development, vanilla JavaScript and HTML and CSS were used.

The CEKG tool offers several features for discovering various types of care pathways that integrate both connecting entities and linking terminology:

\begin{enumerate}
    \renewcommand{\labelenumi}{\textbf{C\arabic{enumi}}}
\item Independent graphs for each patient without consolidating patient activities. \label{C1}
\item Combined graphs for patients without consolidating patient activities. \label{C2}
\item Consolidated patient activities to identify repeated activities for patients with the same multi-morbidity or for specific disorders. \label{C3}
\item Consolidated patient activities to determine how frequently activities related to the treatment of each disorder are repeated for a group of patients with the same multi-morbidity. \label{C4}
\item Consolidated Patient Care Pathways to identify the most frequently repeated activities in the treatment of a group of patients with the same multi-morbidity. \label{C5}
\item Care Pathways that indicate which disorders are treated, untreated, or newly discovered in each admission. \label{C6}
\end{enumerate}

Additionally, we can determine whether to include properties of activities in the graph. For example, should the graph only indicate that a specific clinical test, such as the ABG test, was conducted, or should it also include the test results (e.g., the values of Oxygen, Hemoglobin, ...). Furthermore, we can segment the graph by relating it to the domain or scope of activities.

We provide an online version of the developed tool\footnote{\url{https://cekg-db1cc0d27386.herokuapp.com/}} and an open-source repository\footnote{\url{https://github.com/mnaeimaei/ClinicalEventKnowledgeGraphs_Web/tree/main/Backend_Scripts}} for the back-end scripts. Additionally, A tutorial video\footnote{\url{https://drive.google.com/file/d/1vik3K8XJ1LV5xqdb-wwtv3IUQr1nM3rG/view?usp=sharing}} explaining how to use the tool has also been prepared, along with a complete written tutorial\footnote{\url{https://github.com/mnaeimaei/ClinicalEventKnowledgeGraphs_Web/blob/main/README.md}}.

The tool is designed to handle large datasets, with the primary dataset extracted from the entire MIMIC-IV database. However, it can also be used with alternative datasets. To facilitate testing and provide a template for users to create their own datasets, a test dataset\footnote{\url{https://github.com/mnaeimaei/ClinicalEventKnowledgeGraphs_Web/tree/main/Dataset}} was prepared by extracting a portion of the MIMIC-IV data. The activity titles and patient IDs were modified for de-identification. This test dataset includes essential ICD codes and SNOMED-CT IDs, although the tool is capable of processing the full range of ICD codes and SNOMED-CT ID databases. For convenience, all data in the test dataset was consolidated into a single spreadsheet. However, for practical use, separate CSV files can be imported into the tool.

For building the clinical event knowledge graph, different steps were defined:

\begin{enumerate}
    \item Creating a Labeled Property Graph for the Event Log 
    \item Creating a Labeled Property Graph for Entities Attributes
    \item Creating Relationships Between Entities and Entities Attributes
    \item Creating a Labeled Property Graph for Activity Attributes
    \item Creating a Labeled Property Graph for Activities Domains
    \item Creating a Labeled Property Graph for ICD Codes
    \item Creating Nodes for SNOMED-CT Concepts
    \item Creating Relationships Between Nodes of SNOMED-CT Concepts
    \item Connecting Diagnosed Disorders to ICD Codes
    \item Connecting ICD Codes to SNOMED-CT ID
    \item Connecting Activities to SNOMED-CT Codes
    \item Connecting Activities to Domains and Connecting Domains to SNOMED-CT Codes
    \item Connecting Events to Disorders and Defining Diagnosed Disorders as New Entities
    \item Creating Directly Follows (DF) Relationships and Finalizing the Clinical Event Knowledge Graph
    \item Discovering Care Pathways from the Clinical Event Knowledge Graph
\end{enumerate}

At each step, Neo4j queries are automatically generated based on the input dataset to create graphs or establish relationships between two graphs. These queries will adapt if the dataset is changed. Some queries are designed to clear the database, remove or create constraints, create or modify nodes, or establish relationships between nodes. Since the tool sends the queries to the Neo4j Aura Database, it is possible to view the final clinical event knowledge graph within Neo4j Aura. However, the tool also facilitates the creation of the clinical event knowledge graph offline in a local Neo4J instance by providing all the necessary queries for the user.

\section{Use cases overview}
\label{sec:case}

In this section, we validated the CEKG tool with a case based on the MIMIC-IV dataset~\cite{johnson2023mimic}. 
Two patients with multi-morbidity were considered, including only two entities: \textit{PATIENT} and \textit{ADMISSION}. We used the tools to discover care pathways, denoted as C\ref{C2} and C\ref{C3}, as examples of the care pathways that we can identify. By using the tools, we not only discovered another entity, \textit{Disorder}, but also connected all its activities and entities to ICD-10 and SNOMED-CT to facilitate standardized analysis.

\begin{figure}[tb]
    \centering
    \includegraphics[page=1, scale=0.28, trim=0cm 0cm 125cm 0cm, clip]{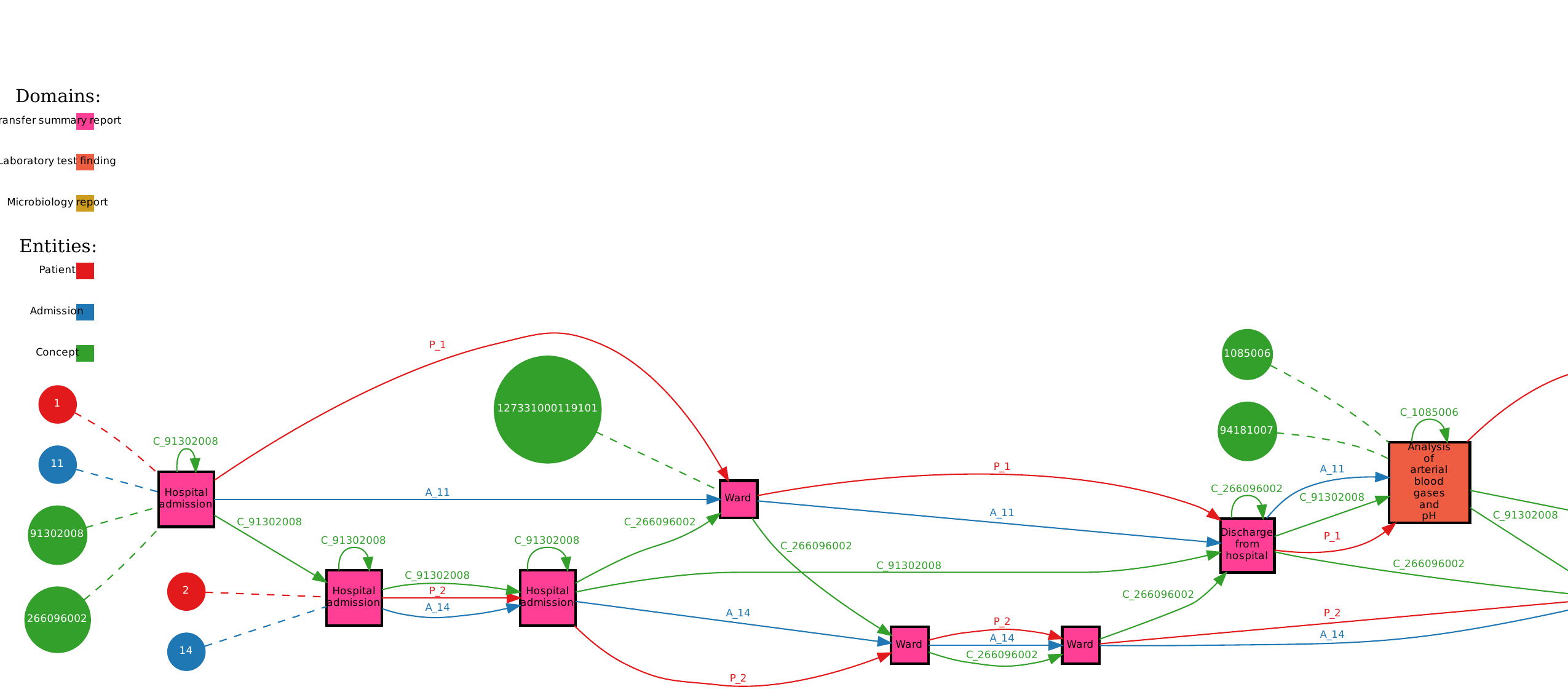}\newline
    \caption{The resulting event graph includes disorder as a new entity. Activities and processes are categorized in a standardized manner, making them interpretable across different healthcare systems.}
    \label{case_study}  
\end{figure}

The C\ref{C2} care pathway discovered frommanner the tool, as shown in Fig.~\ref{case_study}, is the dependent care pathways of two multi-morbid patients, consisting of three entities: \textit{PATIENT} with red circles, \textit{ADMISSION} with blue circles, and \textit{Disorder} with green circles. All activities in the process are mapped to concepts from SNOMED-CT. Additionally, the domains of activities are shown with different colors in the graph. With this type of care pathway, we can determine which activities that happened for these patients are related to which disorders. For example, the graph "Analysis of Arterial blood gases and pH" relates to two disorders with SNOMED-CT 1085006 and 94181007. Furthermore, we can categorize the activities using SNOMED-CT concepts.

\begin{figure}[!ht]
    \centering
    \includegraphics[page=1, scale=0.5, trim=0cm 0cm 0cm 0cm, clip]{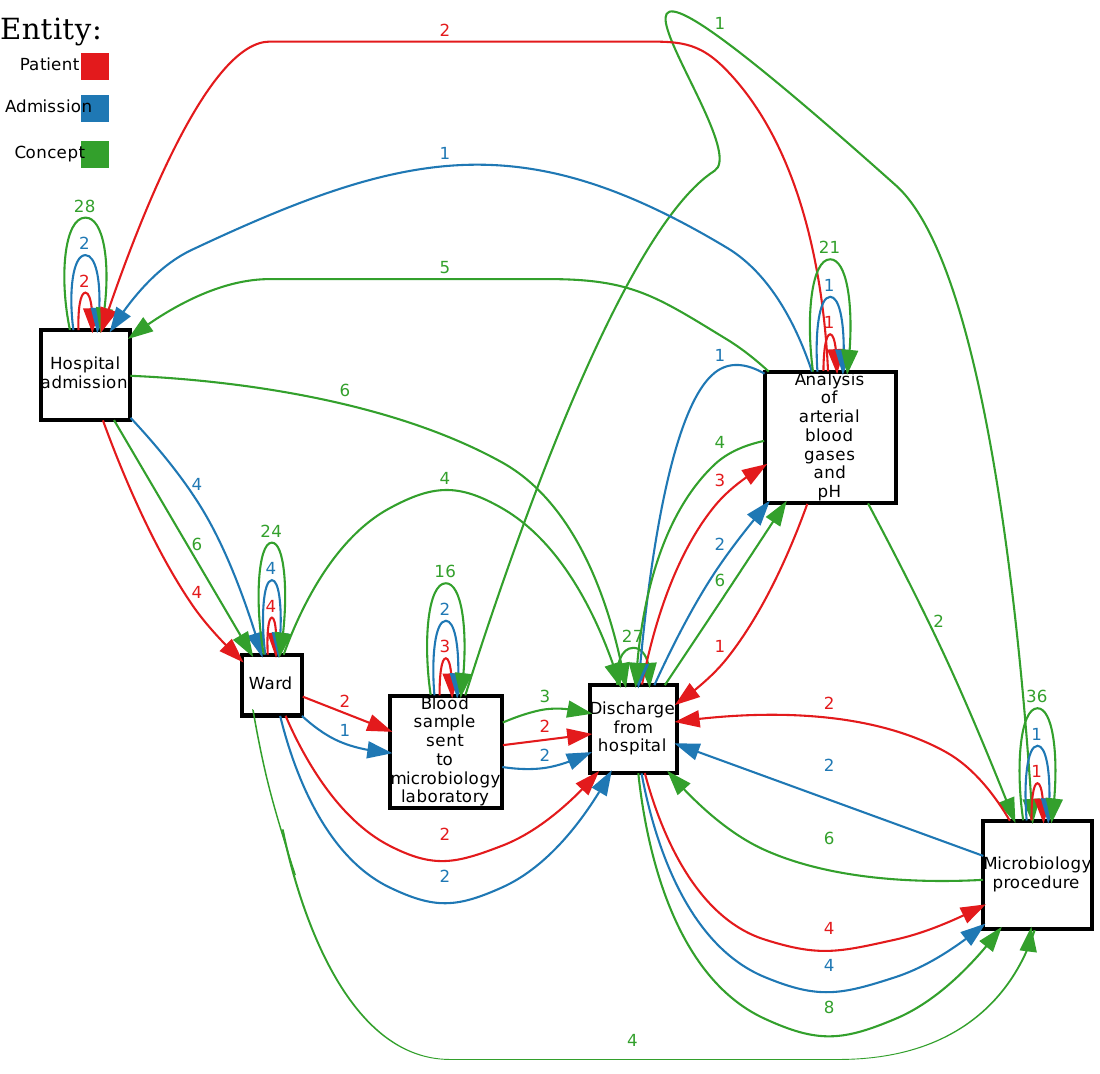}\newline
    \caption{The resulting event graph identifies the most frequently repeated activities in treating two patients, all mapped to SNOMED-CT for standardized analysis.}
    \label{case_study2}  
\end{figure}

The C\ref{C3}, as shown in Fig.~\ref{case_study2}, is the identification of the most frequent activities in the treatment of two patients. For example, we can find out how many times the "Microbiology Procedure" happened after the "Analysis of Arterial Blood Gases and pH" for these two patients. Using SNOMED-CT concepts as a label of activities facilitates the interpretation of the resulting care pathways universally across all health organizations. 

To sum up, the tool streamlines the creation of standardized care pathways by integrating any event log with ICD codes and SNOMED CTs using a graph database. It also automates the generation and execution of the necessary queries for building the graph database, ensuring a seamless process. One area of future research could focus on identifying additional care pathways from the clinical event knowledge graph.


\begin{thebibliography}{8}

\bibitem{marengoni2011aging}
Marengoni, Alessandra, et al. "Aging with multimorbidity: a systematic review of the literature." Ageing research reviews 10.4 (2011): 430-439.

\bibitem{munoz2022process}
Munoz-Gama, Jorge, et al. "Process mining for healthcare: Characteristics and challenges." Journal of Biomedical Informatics 127 (2022): 103994.

\bibitem{naeimaei2023clinical}
Naeimaei Aali, Milad, Felix Mannhardt, and Pieter Jelle Toussaint. "Clinical Event Knowledge Graphs: Enriching Healthcare Event Data with Entities and Clinical Concepts-Research Paper." International Conference on Process Mining. Cham: Springer Nature Switzerland, 2023.

\bibitem{swevels2023object}
Swevels, Ava, Eva L. Klijn, and Dirk Fahland. "Object-Centric Process Mining (and more) using a Graph-Based Approach with PromG." ICPM Doctoral Consortium/Demo. 2023.

\bibitem{johnson2023mimic}
Johnson, Alistair EW, et al. "MIMIC-IV, a freely accessible electronic health record dataset." Scientific data 10.1 (2023): 1.

\end{thebibliography}
\end{document}